\documentclass[10pt]{article}

\usepackage{amsmath}

\usepackage{array}

\usepackage{appendix}

\usepackage{tocloft}                   % Adds Part structure to table of contents

\usepackage{graphicx}

\usepackage{amsfonts}

\usepackage{amssymb}

\usepackage{mathrsfs}

\usepackage{yfonts}

\usepackage{euscript}

\usepackage{centernot}                 % Permits me to trivially make a centred not version of any object

\usepackage{ifsym}                     % Gives a nicer sharp

\usepackage{upgreek}

\usepackage{mathtools}

\usepackage{color}

\usepackage{slantsc}
\usepackage{calligra}

\usepackage{bbold}          % Gives mathbb acting on lower case and, hopefully, numbers...

\usepackage[T1]{fontenc}

\usepackage{epsf}

\usepackage{latexsym}

\usepackage{tipa}

\usepackage{makeidx}

\makeindex

\def\b{\begin{equation}}

\def\e{\begin{equation}}

\def\be{\begin{equation}}              % Longer older ones kept for rext import compatibility.

\def\ee{\end{equation}}

\def\beq{\begin{equation}}

\def\eeq{\end{equation}}

\def\bea{\begin{eqnarray}}

\def\eea{\end{eqnarray}}

\def\m{\mbox{ }}

\def\mma {\m , \m \m }

\def\!{\hspace{-1.6667em}}

\def\c{\cite}

\def\n{\noindent}

\def\u{\underline}

\def\w{\widetilde}

\def\disjoint{\mbox{\scriptsize $\coprod$}}

\def\mC{\mbox{C}}                        % Collinear configuration/ equator in triangleland, can be higher-d regions in other models

\def\mp{\mbox{p}}

\def\bsigma{\mbox{\boldmath$\sigma$}}                   % 

\def\sp{\mbox{\scriptsize p}}

\def\sq{\mbox{\scriptsize q}}

\def\sC{\mbox{\scriptsize C}}

\def\sF{\mbox{\scriptsize F}}

\def\sN{\mbox{\scriptsize N}}

\def\sumi2{\sum\mbox{}_{\mbox{}_{\mbox{\scriptsize $i$=1}}}^2}

\def\sumi3{\sum\mbox{}_{\mbox{}_{\mbox{\scriptsize $i$=1}}}^3}

\def\sumABcycles3{\sum\mbox{}_{\mbox{}_{\mbox{\scriptsize cycles $A,B$=1}}}^{3}}

\def\sumCDcycles3{\sum\mbox{}_{\mbox{}_{\mbox{\scriptsize cycles $C,D$=1}}}^{3}}

\def\sumj3{\sum\mbox{}_{\mbox{}_{\mbox{\scriptsize $j$=1}}}^3}

\def\sumk3{\sum\mbox{}_{\mbox{}_{\mbox{\scriptsize $k$=1}}}^3}

\def\prodiA1{\prod\mbox{}_{\mbox{}_{\mbox{\scriptsize $i$=1}}}^{A - 1}}

\def\bigtimes{\mbox{\Large $\times$}}

\def\es{\m = \m}

\def\:={\m := \m}

\def\=:{\m =: \m}

\def\FrC{\mbox{$\mathfrak{C}$}}                                
                                                       
\def\FrX{\mathfrak{X}}                                         % A set; 
\def\FrS{\mbox{\Large $\mathfrak{s}$}}                         % For the general space, roughly meaning 'in general equipped set'.  
\def\FrU{\mbox{$\mathfrak{U}$}}                                % Open set, with $\{\FrU_{\sfC}\}$ then being an open cover.  
\def\FrV{\mbox{$\mathfrak{V}$}}                                % Another open set, for when two are needed: maps, refinements...
\def\FrW{\mbox{$\mathfrak{W}$}}                                % Yet another open set, for when three are needed for sheaves
\def\FrM{\mbox{$\mathfrak{M}$}}                                % The general manifold 
\def\FrN{\mbox{$\mathfrak{N}$}}                                % A second general manifold, for when needed.
\def\lFrg{\mbox{\Large$\mathfrak{g}$}}                         % Irrelevant group, Lie group.
\def\FrK{\mathfrak{K}}                                         % Another subgroup, when two such are needed.
\def\FrN{\mathfrak{N}}                                         % Normal subgroup.  
                                                  
\def\FrB{\mbox{$\mathfrak{B}$}}                                % Base space of a fibre bundle. 

\def\Hilb{\mbox{{\boldmath$\mathfrak{H}$}ilb}}                 % Hilbert space
\def\bFrY{\mbox{\boldmath$\mathfrak{Y}$}}  

\def\bFrB{\mbox{\boldmath$\mathfrak{B}$}}                      % B-algebra and B*-algebra.                    

\def\FrQ{\mbox{\Large $\mathfrak{q}$}}                               % Configuration space
\def\Phase{\mbox{{\boldmath$\mathfrak{P}$}hase}}                     % Phase space.

\def\bFrR{\mbox{\boldmath$\mathfrak{R}$}}                            % First letter of RigPhase, also used for Riem etc.  Is also, by itself, a ring.
\def\Rig-Phase{\bFrR\mbox{ig-}\Phase}                                % Rigged Phase Space
                                                                													   
\def\FrP{\mbox{\Large $\mathfrak{p}$}}                                 % Preshape space.     
\def\FrR{\mbox{\boldmath$\mathfrak{R}$}}                             % Relational space

\def\bFrR{\mbox{\boldmath$\mathfrak{R}$}}                            % Used in regularity structure symbol

\def\bFrR{\mbox{\boldmath$\mathfrak{R}$}}                            % Used in incipient regularity structure symbol

\def\1mat{\u{\u{1}}}                                                 % unit-entry matrix

\def\Positive-Modespace{\mbox{{\boldmath$\mathfrak{M}$}odespace$^+$}}% Positive modespace

\def\POSITIVE-MODESPACE{\mbox{{\boldmath$\mathfrak{M}$}ODESPACE$^+$}}% Positive modespace alongside scalar field matter inhomogeneous modes.
                                                                                                                             														
\def\bFrS{\mbox{\Large $\mathfrak{s}$}}                              % For the general space, roughly meaning 'in general equipped set'.  
			
\def\bFrT{\mbox{\boldmath$\mathfrak{T}$}}                            %

\def\FrO{\mbox{$\mathfrak{O}$}}                                      % Individual gauge orbits.

\def\Kin-Hilb{\mbox{{\boldmath$\mathfrak{K}$}in-\Hilb}}                     % Dynamical Hilbert space 

\def\Mid-Hilb{\mbox{{\boldmath$\mathfrak{M}$}id-\Hilb}}                     % Dynamical Hilbert space 

\def\Dyn-Hilb{\mbox{{\boldmath$\mathfrak{D}$}yn-\Hilb}}                     % Dynamical Hilbert space 

\def\5Star{\mbox{\Large$\star$}}              % Rectified time derviative actually used

\begin{document}

\begin{titlepage}

\begin{center}

{\bf\Large Shape Theories. II. Compactness Selection Principles} 

\vspace{0.1in}
\m

{\bf Edward Anderson}$^*$  

\m 

\end{center}

\begin{abstract}

Shape(-and-scale) spaces - configuration spaces for generalized Kendall-type Shape(-and-Scale) Theories - are usually not manifolds but stratified manifolds. 
While in Kendall's own case - similarity shapes - the shape spaces are analytically nice - Hausdorff - 
for the Image Analysis and Computer Vision cases - affine and projective shapes - they are not: merely Kolmogorov.  
We now furthermore characterize these results in terms of whether one is staying within, 
or straying outside of, some compactness conditions which provide protection for nice analytic behaviour.  
We furthermore list which of the recent wealth of proposed shape theories lie within these topological-level selection principles for technical tractability. 
Most cases are {\sl not} protected, by which the merely-Kolmogorov behaviour may be endemic and 
the range of technically tractable Shape(-and-Scale) Theories very limited. 
This is the second of two great bounds on Shape(-and-Scale) Theories, each of which moreover have major implications for Comparative Background Independence as per Article III. 

\end{abstract}

\m 

\n Mathematics keywords: Applied Topology: stratified manifolds, Hausdorff versus Kolmogorov separation. 
Used for Shape Theory (underlying Shape Statistics and the below).

\m

\n Physics keywords: Configuration spaces. Image spaces (Image Analysis, Computer Vision). Relational Mechanics, $N$-Body Problem, Background Independence. 

\m

\n PACS: 04.20.Cv, 02.40.-k 

\m 

\n $^*$ dr.e.anderson.maths.physics *at* protonmail.com 

%====================================================================================================================================================================================
%====================================================================================================================================================================================
\section{Introduction}
%====================================================================================================================================================================================
%====================================================================================================================================================================================

\n We continue our treatise of generalized Kendall (Scale-and-)Shape Theories \cite{Kendall84, Kendall89, Kendall, FileR, 
                                                                                 Kendall87, ASphe, 
																				 Sparr, MP03, GT09, 
																				 MP05, Bhatta, AMech, DM16, PE16, KKH16,   
																				 ATorus, A-Mink, Minimal-N} 
by placing new topological-level selection principles for technical tractability among such theories.  

\m 

\n Let us first note that, within the nongeneric case of manifolds possessing generalized Killing vectors as per Article I \cite{A-Killing}, 
the corresponding shape(-and-scale) spaces, alias {\it relational spaces}, are usually stratified manifolds rather than manifolds. 
These are still topological spaces, so we start in Sec 2 with a brief outline of these and their properties as used in the current Article. 
This goes as far as describing how topological manifolds are a specialization of topological spaces which serves to model carrier spaces
\be 
\FrC^d = \FrM
\label{Carrier}
\ee  
$N$-point constellation spaces 
\be 
\FrQ(d, N) \:= \bigtimes_{I = 1}^N \FrM
\label{Constell}
\ee 
 and an exceptional few relational spaces, 
including Kendall's own Shape Theory's \cite{Kendall84, Kendall89, Kendall, FileR, I-II-III, Minimal-N} shape spaces in 1- and 2-$d$.  

\m 

\n Quotients of constellation space, by some automorphism group 
\be 
\lFrg = Aut(\FrM, \bsigma)
\label{Aut}
\ee 
preserving geometrical level of structure $\bsigma$, are known as {\it relational spaces}  
\be 
\FrR(d, N; \lFrg) \es \frac{\FrQ(\FrM, N)}{\lFrg} \es \frac{\FrQ(d, N)}{Aut(\FrM, \bsigma)}   \m . 
\label{Rel}
\ee 
These are in general stratified manifolds, i.e.\ a somewhat more general kind of topological spaces (Sec 2.1) 
                                                                         than topological manifolds (Sec 2.2), 
																  which are themselves introduced in Sec 3.  
Three strategies for handling these are outlined in Sec 4, revisited in Sec 7 from the generic point of view, and in Sec 10's comparative account.      
Sec 6 furthermore considers the breakdown of applicability of fibre bundles in the stratified manifold setting, 
including (pre)sheaf and differential space replacements.  

\end{titlepage}

\n To date, relational stratified manifolds have moreover been observed to occur in three bands as regards topological separation properties (Sec 5). 

\m 

\n {\bf Type 1)} Hausdorff-separated stratified manifolds. 
These occur e.g.\ in Kendall's own Similarity Shape Theory \cite{Kendall84, Kendall89, Kendall} in $\geq 3$-$d$, starting with the pure-shape 3-$d$ 3-Body Problem. 
This gives the open hemisphere adjoined to a stratificationally distinct circular-equator edge of collinear shapes.  

\m 

\n {\bf Type 2)} Hausdorff stratified manifolds modulo removal of isolated bad points, such as the usual (scaled, Euclidean $d \geq 2$) 
$N$-Body Problem's maximal coincidence-or-collision O.\footnote{This a portmanteau of physical maximal collision of particles 
%OOOOOOOOOOOOOOOOOOOOOOOOOOOOOOOOOOOOOOOOOOOOOOOOOOOOOOOOOOOOOOOOOOOOOOOOOOOOOOOOOOOOOOOOOOOOOOOOOOOOOOOOOOOOOOOOOOOOOOOOOOOOOOOOOOOOOOOOOOOOOOOOOOOOOOOOOOOOOOOOOOOOOOOOOOOOOOOOOOO
and maximal coincidence of all points, be they geometrical figure vertices or statistical location data points, all piled up at exactly the same point in carrier space.} 
%OOOOOOOOOOOOOOOOOOOOOOOOOOOOOOOOOOOOOOOOOOOOOOOOOOOOOOOOOOOOOOOOOOOOOOOOOOOOOOOOOOOOOOOOOOOOOOOOOOOOOOOOOOOOOOOOOOOOOOOOOOOOOOOOOOOOOOOOOOOOOOOOOOOOOOOOOOOOOOOOOOOOOOOOOOOOOOOOOOO

\m 

\n {\bf Type 3)} Merely Kolmogorov-separated stratified manifolds. 
In this regard, we mention in particular Affine \cite{Sparr, MP03} and Projective \cite{MP05, PE16} Shape Theories, 
motivated as view-from-infinity and view-from-an-arbitrary-vantage-point respectively in each of Image Analysis and Computer Vision. 
These shape theories, originally posited by mathematican Gunnar Sparr and by statisticians Kanti Mardia and Victor Patrangenaru respectively, 
have since been shown to exhibit this type of stratification in \cite{GT09} and \cite{KKH16} respectively.  

\m

\n The current Article serves to explain this banding in terms of underlying topological selection principles (Sec 8).  
Our starting point is to notice that Kendall's own similarity shape spaces -- of the most benevolent type 1) -- can be expressed as 'quotients of a 
compact space by a compact Lie group': of his preshape spheres by the rotations. 
While one can keep on declaring carrier spaces (\ref{Carrier}), 
formulating constellation spaces as finite product spaces of these (\ref{Constell}), 
and then quotienting them by whatever geometrical automorphism groups (\ref{Aut}) the poset of generalized Killing equations on $\FrM$ supports, 
these two additional compactness niceties for Kendall's Shape Theory turn out to seldom extend. 
We argue moreover that much of the individual success of Kendall's Shape Theory is due to these two compactness niceties.  
More specifically, we show that 'quotient of a compact space' and especially `quotient by a compact Lie group' confer protective theorems (Sec 8). 
In particular, the latter guarantees a locally-compact Hausdorff second-countable (LCHS) quotient: 
an analytically particularly well-behaved type of topological space. 
This result is, among other things, a selection principle for stratified manifolds that are guaranteed to enjoy Type 1)'s Hausdorffness. 
LCHS is moreover the class of stratified manifolds studied, separately, by mathematicians Matthias Kreck \cite{Kreck} and Jedrzej \'{S}niatycki \cite{LS08, SniBook}. 
Due to this, we also point to subcartiesian differential space results 
-- \'{S}niatycki's approach, building on much earlier work of mathematician Roman Sikorski \cite{Sikorski} --  
being applicable to (shape-and-)scale spaces in Sec 8.7.  

\m

\n Sec 9 compares the current Article with that of Kendall et al on the topology of similarity shape space, 
whereas Sec 10 serves to comment on \c{GT09}'s Mumfordian stance and the remit of its applicability.  
 
\m  

\n On the one hand, our compactness characterization motivates some further Shape Theories which share these features, starting with 
isometric Shape-and-Scale Theory on circles, tori, spheres and real-projective spaces.
On the other hand, most other Relational Theories considered to date - while well-motivated - 
fail one or both of our protective compactness criteria for technical tractability, 
with many involving in particular an irreducible quotient by a noncompact Lie group.
In particular, Projective and Conformal Shape Theories on flat space, and Spacetime-events Shape Theories, 
fail both, while Affine Shape theory and Equi-top-form Scaled-Shape Theory involve quotienting the preshape sphere by noncompact Lie groups.
We finally point to most cases {\sl not} being protected, 
by which the merely-Kolmogorov behaviour may be endemic and the range of technically tractable (Scale-and-)Shape Theories very limited. 
This has consequences for the Comparative Theory of Background Independence as per Article III.

%====================================================================================================================================================================================
%====================================================================================================================================================================================
\section{Topological spaces and topological manifolds}\label{Top-S-M}
%====================================================================================================================================================================================
%====================================================================================================================================================================================

\n{\bf Global Problem 1} Configuration space reductions 
\be 
\FrQ \m \longrightarrow \m \widetilde{\FrQ}   \es  \frac{\FrQ}{\lFrg}
\ee 
{\sl usually kick one out of the class of manifolds into the class of stratified manifolds}.  
In the current study of Relational Theories, this means that the relational spaces -- shape(-and-scale) spaces -- are in general stratified manifolds.

\m 

\n Since stratified manifolds are rather unfamiliar outside of Geometry, Dynamical Systems and Theoretical Statistics they require some explaining. 
A first point of order is that, while these are not topological {\sl manifolds}, they are still topological {\sl spaces}. 
We thus start by recollecting this standard albeit rather less structured notion.

%===================================================================================================================================================================================
\subsection{Some basic topological space notions} 
%===================================================================================================================================================================================

\n{\bf Definition 1} A {\it topological space} \cite{Sutherland, Munkres, Willard, Engelking, Nagata} is a set $\FrX$ alongside a collection 
\be 
\bFrT = \left\{ \mbox{open } \m \FrU_i \subseteq \FrX \, | \, i \in {\cal I} \, : \m \mbox{ an arbitary index set } \right\}  
\ee 
with the following properties. 

\m 

\n{\bf Topological Spaces 1)} $\FrX, \emptyset \, \in \, \bFrT$.

\m 

\n{\bf Topological Spaces 2)} 
\be 
\bigcup_{j \in {\cal J}} \FrU_j \, \in \,  \bFrT \m \mbox{ (closure under arbitrary unions) }  \m . 
\ee 
\n{\bf Topological Spaces 3)} 
\be 
\bigcap_{k = 1}^K \FrU_j \, \in \, \bFrT \m \mbox{ (closure under finite intesections) }  \m . 
\ee 
\n Some topological properties that the current Article makes use of are as follows. 

\m

\n{\bf Definition 2} A collection of open sets $\{ \FrU_{l}, l \in {\cal L} \}$ is an {\it open cover} \c{Sutherland, Lee1} for $\FrX$ if 
\be 
\FrX = \bigcup_{ l \in {\cal L} } \FrU_{l} \m .
\ee 
\n{\bf Definition 3} A subcollection of an open cover that is itself still an open cover is termed a {\it subcover}, 
$\{\FrV_{m}\}$ for $m \in {\cal M} \subseteq {\cal L}$. 
 
\m  

\n{\bf Definition 4} An open cover $\{\FrW_{n}\}$ is a {\it refinement} \cite{Munkres, HY} of $\{\FrU_{l}\}$ if 
\be 
\mbox{ each } \m \FrV_{n} \m \mbox{ has a } \m \FrU_{l} \m \mbox{ such that } \m \FrV_{n} \subset \FrU_{l}    \m .  
\ee 
$\{\FrV_{n}\}$ is furthermore {\it locally finite} if 
\be 
\mbox{ each } \m x \in \FrX \m \mbox{ has an open neighbourhood } \m \FrN_x \m 
\mbox{ such that only finitely many } \m \FrV_{n} \m \mbox{ obey } \m \FrN_x \bigcup \FrV_{n} \neq \emptyset  \m .  
\ee 
\n{\bf Definition 5} A {\it base} \cite{Sutherland, Munkres} for a topological space $\langle \FrX, \bFrT \rangle$ is a subcollection 
\be 
\bFrT \supseteq \bFrB \:= \langle \, \FrB_p \mma i \in {\cal P} \, \rangle 
\ee
such that every open subset $\FrO \in \FrX$ is covered by elements of $\bFrB$, 
\be 
\FrO = \bigcup_{q \in {\cal Q}} \FrB_q   \m . 
\ee   
\n{\bf Definition 6} A {\it local base} \cite{Munkres} $\bFrB_x$ for a topological space $\langle \FrX, \bFrT \rangle$ at point $x \in \FrX$ 
is a collection of neighbourhoods of $x$ such that every neighbourhood $\FrU$ of $x$ contains some $\FrB \in \bFrB_x$.   

\m 

\n{\bf Definition 7} A {\it sub-base} \cite{Willard} for a topological space $\langle \FrX, \bFrT \rangle$ is a subcollection of its subsets, 
\be 
\bFrT \supseteq \bFrS \:= \langle \, \FrS_s \mma s \in {\cal S} \, \rangle 
\ee
such that $\FrX$ alongside the collection of all finite intersections of elements of $\bFrS$ form a base.  

\m 

\n{\bf Definition 8} Let $\FrX$ and $\bFrY$ be topological spaces. 
{\it Homeomorphisms} are continuous maps 
\be 
\phi : \FrX \longrightarrow \bFrY 
\ee 
in possession of continuous inverses $\phi^{-1}$. 

\m 

\n{\bf Definition 9} {\it Topological properties} \c{Armstrong, Willard, Engelking, Nagata} are those attributes of a topological space that are invariant under all homeomorphisms. 

\m

\n{\bf Structure 1} Notions of {\it separation} are topological properties which indeed involve separating two objects 
(points, certain kinds of subsets) by encasing each in a disjoint subset.  
The following is a particular such.

\m 

\n{\bf Definition 10} A topological space is {\it Hausdorff} \cite{Sutherland, Lee1, Willard} if   
$$
\mbox{for } \m \mbox{$x, \, y \in \FrX \mma x \neq y \mma  \exists$ \m open sets \m ${\FrU}_x \mma {\FrU}_y \in \bFrT$}
$$
\beq
\mbox{such that \m $x \in {\FrU}_x \mma y \in {\FrU}_y$ \m and \m ${\FrU}_x \bigcap {\FrU}_y = \emptyset$}          \m . 
\label{Hausdorffness}
\eeq
I.e.\ any pair of points can be separated by open sets.  

\m 

\n{\bf Structure 2} Hausdorffness allows for each point to have a neighbourhood without stopping any other point from having one.
This generalizes a property of $\mathbb{R}$ that much Analysis depends upon; Hausdorffness guarantees in particular in this way uniqueness for limits of sequences. 

\m 

\n{\bf Definition 11} $\langle \FrX, \bFrT \rangle$ is {\it first countable} \cite{Munkres, Willard} if each $x \in \FrX$ has a countable local base. 

\m

\n{\bf Definition 12} $\langle \FrX, \bFrT \rangle$ is {\it second countable} \cite{Munkres, Willard} if it admits a countable base. 

\m 

\n{\bf Remark 1} Second countability is a more stringent global condition to first countability's local condition.  

\m 

\n{\bf Remark 2} These countability criteria protect one's topology from containing `too many' open subsets. 
On the one hand, first countability corresponds to topological spaces in which sequences suffice to detect most topological properties. 
On the other hand, second-countability has the additional useful feature of guaranteeing equivalence of compactness and sequential compactness \cite{Willard}.

\m  

\n{\bf Definition 13} $\langle \FrX, \bFrT \rangle$ is {\it locally Euclidean (LE)} \c{Lee1} if every point $x \in \FrX$ has a neighbourhood $\FrN_x$ 
that is homeomorphic to $\mathbb{R}^p$: Euclidean space. 

\m 

\n{\bf Definition 14} $\langle \FrX, \bFrT \rangle$ is {\it compact} \cite{Armstrong, Sutherland, Lee1, Willard} 
\be 
\mbox{if every open cover of } \m \FrX \m \mbox{ has a finite subcover }                                      \m .  
\ee  
\n{\bf Remark 3} Compactness is useful e.g.\ through its generalizing continuous functions being closed and bounded on a closed interval of $\mathbb{R}$ 
(i.e.\ the Heine--Borel Theorem \cite{Sutherland, Munkres}).

\m

\n{\bf Definition 15} $\langle \FrX, \bFrT \rangle$ is {\it paracompact} \cite{Lee1, Munkres, Willard} 
\be 
\mbox{if every open cover of } \m \FrX \m \mbox{ has a locally finite refinement}                             \m . 
\ee 
\n{\bf Remark 4} Paracompactness is useful e.g.\ via Hausdorff paracompact (HP) spaces permitting use of partitions of unity \cite{Lee2, Munkres, Nagata}.

%===================================================================================================================================================================================
\subsection{Topological manifolds} 
%===================================================================================================================================================================================

\n{\bf Definition 16} {\it Topological manifolds} \cite{Lee1} are topological spaces blessed with the combination of being Hausdorff, second-countable and locally Euclidean (LEHS).  

\m  

\n{\bf Remark 5} Hausdorffness and second-countability is a useful combination -- in the manner of a balance point -- as regards analytical tractability.
The preceding subsection's comments about use of sequences fit together to render 
topological spaces which are not Hausdorff as having too few open sets for much of Analysis, whereas 
topological spaces which are not second-countable have too many. 

\m 

\n{\bf Remark 6}  Local Euclideanness moreover underlies the use of charts in subsequent study of differentiable manifolds, 
i.e.\ topological manifolds further equipped with differentiable structure. 

\m 

\n{\bf Remark 7}  Manifolds' defining topological properties imply paracompactness as well \cite{Munkres, Willard, Lee2}.  
The partitions of unity thus guaranteed then readily permit integral Calculus thereupon.

%==================================================================================================================================================================
\subsection{Productive topological properties}
%==================================================================================================================================================================

\n{\bf Structure 3} product of topological spaces, and the product topology 

\m 

\n{\bf Definition 17} A topological property is {\it productive} if it is preserved under products of topological spaces.

\m 

\n{\bf Proposition 1} i) Hausdorffness, ii) second-countability, and iii) local-Euclideanness are productive. 

\m 

\n{\bf Remark 8} `Productive' in the often simpler finite sense suffices for the purposes of the current Article, as is clear from our main application 
of this Proposition below. 

\m 

\n{\bf Corollary 1} Manifoldness is productive. 

\m 

\n{\bf Corollary 2} If carrier space $\FrC^d$ is a manifold, the corresponding constellation spaces $\FrQ(\FrC^d, N)$ are manifolds.

%==================================================================================================================================================================
\subsection{Quotientive topological properties}
%==================================================================================================================================================================

\n{\bf Remark 9} Recollect moreover that nontrivial relational spaces are quotients of constellation spaces by automorphism groups. 
Mathematically, this is a subcase of \cite{Munkres, Lee1}
\be 
\frac{\mbox{(manifold)}}{\mbox{(Lie group)}}   \es  \frac{\FrM}{\lFrg} \m .
\ee
In this way, quotient spaces enter our study.

\m 

\n{\bf Structure 4} Quotienting a topological space according to an equivalence relation, $\langle \FrX, \bFrT\rangle/\,\widetilde{\mbox{ }}$, 
produces the corresponding {\it quotient topology} \cite{Munkres, Lee1}.

\m 

\n{\bf Definition 18} A topological property is {\it quotientive} if it is preserved under quotients. 

\m 

\n{\bf Remark 10} This unfortunately seldom occurs. In particular, we have the following result.  

\m 

\n{\bf Proposition 2} \cite{Munkres, Lee2, Willard} i) Hausdorffness, ii) second-countability, and iii) local-Euclideanness are not in general quotientive.

\m 

\n{\bf Corollary 3} Manifoldness is not in general quotientive.

\m 

\n{\bf Corollary 4} While constellation spaces $\FrQ(\FrC^d, N)$ are manifolds, relational spaces ${\cal R}(\FrC^d, N; \lFrg)$
-- quotients of these by geometrical automorphism Lie groups -- are not in general manifolds.

%====================================================================================================================================================================================
%====================================================================================================================================================================================
\section{Stratified manifolds}\label{Strata}
%====================================================================================================================================================================================
%====================================================================================================================================================================================

Stratified manifolds are a relatively well-behaved class of spaces arising by loss of the locally-Euclidean bastion of manifoldness, 
as often occurs e.g.\ in quotienting manifolds. 
This loss corresponds to a stratified manifold in general having {\sl multiple} dimensions, in a piecewise manner.
Stratified manifolds moreover entail but a {\sl local} loss of local Euclideanness.
They furthermore have some fairly benevolent rules for `patching together' a stratified manifold's pieces of different dimensionality, 
as put forward by mathematicians Hassler Whitney and Ren\'{e} Thom \cite{W-T, Whitney65, Thom69}. 

\m 

\n{\bf Structure 1} Let $\FrX$ be a topological space that can be split according to 
\be
\FrX \es \FrX_{\sp} \cup \FrX_{\sq}                                                                                                          \m .
\ee 
for which 
\be 
\mbox{dim}_{\sp}(\FrX) = \mbox{dim}(\FrX)
\ee 
and 
\be 
\FrX_{\sq} \:= \FrX - \FrX_{\sp}                                                                                                            \m .
\ee  
Next consider recursive such splittings, so e.g.\ $\FrX_{\sq}$ further splits into $\{\FrX_{\sq}\}_{\sp}$ and $\{\FrX_{\sq}\}_{\sq}$.
Then set
\be 
\FrM_1 = \FrX_{\sp} \mma \FrM_2 = \{\FrX_{\sq}\}_{\sp} \mma \FrM_3 = \{\{\FrX_{\sq}\}_{\sq}\}_{\sp}  \m . \m . \m .
\ee 
to obtain  
\be 
\FrX = \FrM_1 \cup \FrM_2 \cup  \, ... \,  , \mbox{dim}(\FrX) = \mbox{dim}(\FrM_1) \, > \, \mbox{dim}(\FrM_2) \, > \m . \m . \m .           \m ,
\ee
where each $\FrM_{I}$, $I = 1, 2, \, ...$ is itself a manifold.
This provides a partition of $\FrX$ by dimension. 
$\FrX$ is moreover only a topological manifold in the case of a trivial (i.e.\ single-piece) partition.

\m 

\n{\bf Definition 1} A {\it strict} partition of a topological space is a (locally finite) partition into strict manifolds. 
A manifold $\FrM$ within a $m$-dimensional open set $\FrW$ is $\FrW${\it -strict} if its $\FrW$-{\it closure} 
\be
\overline{\FrM}  \:=  \FrW - \mbox{Clos}\,\FrM
\ee 
and the $\FrW$-{\it frontier} 
\be 
\FrM^{\sF}       \:=  \overline{\FrM} - \FrM
\ee 
are topological spaces in $\FrW$. 

\m 

\n{\bf Definition 2} A set of manifolds in $\FrW$ has the {\it frontier property} if, for any two distinct such, say $\FrM$ and $\FrM^{\prime}$,  
\beq
\mbox{ if }   \m  \FrM^{\prime} \cap \overline{\FrM} \m \neq \m  \emptyset \mma 
\mbox{ then } \m  \FrM^{\prime} \subset \overline{\FrM}              \m \mbox{ and } \m \m 
                  \mbox{dim}(\FrM^{\prime}) \, <  \, \mbox{dim}({\FrM}) \mbox{ } .
\label{Frontier}
\eeq
A partition into manifolds itself has the frontier property if the corresponding set of manifolds does.  

\m 

\n{\bf Definition 3} A {\it stratification} of $\FrX$ \cite{Whitney65} is a strict partition of $\FrX$ that possesses the frontier property. 
The corresponding set of manifolds are known as the {\it strata} of the partition.

\m

\n{\bf Definition 4} The {\it principal stratum} is the one whose corresponding {\sl orbit} is of minimal dimension.

\m 

\n{\bf Remark 1} Stratified manifolds have additionally been equipped with differentiable structure \cite{Thom69}; 
see e.g.\ \cite{Pflaum, SniBook} for more modern accounts.

%===================================================================================================================================================================================
%===================================================================================================================================================================================
\section{Strategies for handling stratification} 
%===================================================================================================================================================================================
%===================================================================================================================================================================================

\n{\bf Strategy A)} {\it Excise Strata}. Discard all bar the principal stratum \cite{Marsden, BF}.

\m 

\n{\bf Strategy B)} {\it Unfold Strata}. Unfold non-principal strata, so that these end up possessing the same dimension as the principal stratum \cite{Fischer86}.  

\m

\n{\bf Strategy C)} {\it Accept All Strata} \cite{AConfig, ABook}. 

\m 

\n There are moreover the following contextual differences. 

\m 

\n{\bf Context 1)} Non-principal strata are `not meaningful to the underlying system being modelled' 
(the physical setting's subcase of this expression is `unphysical'). 

\m 

\n{\bf Context 2)} Non-principal strata are meaningful to the underlying system being modelled.

\m 

\n{\bf Context 2.1)} Non-principal strata's distinctions in dimension are meaningful to the underlying system being modelled.   

\m 

\n{\bf Context 2.2)} It is a non-principal stratum's unfolded dimension, rather, which is meaningful to the underlying system being modelled. 

\m 

\n{\bf Remark 1} Excise Strata matches context 1, but is in conflict with context 2, in which apart from violating the modelling, it can be a crude approximation.
Unfold Strata is in conflict with contexts 1 and 2.1, but matches context 2.2.  
Accept All Strata is not necessary in context 1, but matches context 2.1.  

\m 

\n{\bf Remark 2} Background Independence requires one's strategy about strata to match its modelling context
(e.g.\ this can be thought of as quite an advanced example of Leibniz's \cite{L} Identity of Indiscernibles).
Unfold Strata moreover has the further complications of needing to be meaningful to the modelling situation 
                                                                                and independent of mathematical details of the unfolding procedure.  
The current Article's analysis of strategies for stratification is the most advanced to date, superceding \cite{ABook}'s.

\m 

\n{\bf Remark 3} On the one hand, Excise Strata and Unfold Strata 
keep one within the familiar and mathematically tractable remit of Manifold Geometry and Fibre Bundles thereover.
On the other hand, Accept all Strata left harder mathematics being required (this and the next two sections).

%===================================================================================================================================================================================
%===================================================================================================================================================================================
\section{Analytic tractability classification of Relational Theory's stratified manifolds} 
%===================================================================================================================================================================================
%===================================================================================================================================================================================

\n The following classification of stratified manifolds is useful in considering relational spaces.

%===================================================================================================================================================================================
\subsection{Type 1) ``analytically nice' stratified manifolds} 
%===================================================================================================================================================================================

These keep one of the three bastions of manifoldness -- Hausdorffness -- and more-or-less keep another -- second countability -- as detailed below, 
and moreover supplement these with the following further analytic nicety.

\m 

\n{\bf Definition 1} A topological space $\FrX$ is {\it locally compact (LC)} \cite{Lee1} if each point $\mp \in \FrX$ is contained in a compact neighbourhood.    

\m

\n{\bf Remark 1} Local compactness aids Analysis by sufficing to enable notions of compactness, 
from locally attaining bounds through to compact generation and 1-point compactification.

\m  

\n{\bf Structure 1} Our first class of `analytically nice spaces' is thus {\it LCHS spaces} \cite{Lee1, Lee2, Willard}: locally-compact Hausdorff second-countable.

\m 

\n{\bf Remark 2} These have further analytical niceties; in particular LCH spaces behave similarly to complete metric spaces,  
and LCHS spaces furthermore admit exhaustion by compact sets.  

\m 

\n{\bf Remark 3} All manifolds are locally-compact \cite{Lee2}: LEHS $\Rightarrow$ LCHS.  

\m

\n{\bf Structure 2} Our first class of `analytically nice' stratified manifolds consists of the {\it LCHS stratified manifolds}. 
These have moreover been studied by Kreck \cite{Kreck} and by \'{S}niatycki \cite{SniBook}.   
                                                                    
\m 

\n{\bf Structure 3}  {\sl LCHP spaces} \cite{Munkres} -- locally compact Hausdorff paracompact -- are another analytically well-behaved class.

\m 

\n{\bf Remark 4} There is moreover considerable degeneracy between paracompactness and second-countability \cite{Munkres, Harvard-14}, 
by which considering a portmanteau of LCHS and LCHP is rather appropriate.  
\be
\mbox{For LC spaces} \mma \mbox{S $\Rightarrow$ P} \mma \mbox{so LCHS generalizes LCHP} \m .  
\label{LCHSP}
\ee
\n{\bf Structure 4} Our second class of `analytically nice' stratified manifolds is then the {\it LCHP stratified manifolds}. 
This case was moreover studied by mathematician Markus Pflaum \cite{P00, Pflaum}, 
a decade prior to the above treatments of the somewhat more general LCHS stratified manifolds.  

\m

\n{\bf Remark 5} LCHS and LCHP are moreover standard and well studied-packages in their own right, above and beyond these applications to stratified manifolds. 
They are used in many further areas of Mathematics (topological groups, groupoids, Polish spaces, random sets \cite{Kendall74}...)
This common mathematical base moreover gives a sense of {\sl compatibility}.
This is a significant feature in composing Background Independence aspects 
and in inter-relating different approaches to these, or to Shape Theory, or to the Foundations of Geometry. 

\m 

\n{\bf Example 1} Kendall's shape spaces in 1- and 2-$d$ are manifolds, and thus both LCHS and LCHP by Remark 2 above and Remark 10 of Sec 2. 

\m 

\n{\bf Example 2} Kendall's shape spaces in $\geq 3$-$d$ are nontrivially stratified and thus not manifolds; they are moreover LCHS stratified manifolds.  

\m 

\n{\bf Remark 6} We moreover have cause to make the following distinction in our study of stratified manifolds arising as relational spaces.

\m 

\n{\bf Type 2) Stratified manifolds with isolated instances of obstructions to Analysis}.  

\m 

\n{\bf Type 3) Stratified manifolds with widespread obstructions to Analysis}.

%===================================================================================================================================================================
\subsection{Type 2) stratified manifolds with isolated obstructions to Analysis}
%===================================================================================================================================================================

\n{\bf Remark 7} Type 2 refers in particular to singularities occurring at certain stratificationally-distinguished isolated `bad points'.    

\m 

\n{\bf Remark 8} Excision is then a mathematical option.   
Under certain circumstances, this is reasonable as regards the modelling situation as well (e.g.\ physically reasonable).  

\m 

\n{\bf Example 3} For $N \geq 2$, the shape-and-scale space 
\be 
\FrR(d, N) \es \frac{\FrQ(d, N)}{Eucl(d)} 
\ee 
has a maximal coincidence-or-collision O that is arbitrarily near to all shapes.
O is furthermore stratificationally distinguished for $d \geq 2$.
This is since $SO(d)$ is then a nontrivial group, which acts only as $id$ on O's single point, but acts via some nontrivial continuous subgroup on all other shapes.  
O is moreover a bad point if additionally $N \geq 3$.

\m

\n Furthermore \cite{Iwai87, Cones, FileR}, 
\be 
\FrR(d, N) = \mC(\FrS(d, N)) 
\ee 
for $\mC$ the topological- and metric-level cone \cite{Lee1, FileR},  
I.e.\ the Leibnizian relational space corresponding to quotienting out by the Euclidean group is the cone over Kendall's shape space. 
Therein, O plays the role of cone point.
For $d \geq 2, N \geq 3$, this is moreover a singular point.\footnote{In 1-$d$, the cone just returns the flat relative space $\mathbb{R}^{N - 1}$ 
%OOOOOOOOOOOOOOOOOOOOOOOOOOOOOOOOOOOOOOOOOOOOOOOOOOOOOOOOOOOOOOOOOOOOOOOOOOOOOOOOOOOOOOOOOOOOOOOOOOOOOOOOOOOOOOOOOOOOOOOOOOOOOOOOOOOOOOOOOOOOOOOOOOOOOOOOOOOOOOOOOO
as the cone over Kendall's preshape sphere $\mathbb{S}^{N - 2}$.
For $N = 2$, there is only one shape: the finite interval, so just $\mathbb{R}$ -- the cone over a point -- ensues.
In either case, the cone point has no associated topological or geometrical badness.}
%OOOOOOOOOOOOOOOOOOOOOOOOOOOOOOOOOOOOOOOOOOOOOOOOOOOOOOOOOOOOOOOOOOOOOOOOOOOOOOOOOOOOOOOOOOOOOOOOOOOOOOOOOOOOOOOOOOOOOOOOOOOOOOOOOOOOOOOOOOOOOOOOOOOOOOOOOOOOOOOOOO

\m 

\n The maximal coincidence-or-collision O is moreover also unphysical, 
since the notion of a point particle surely breaks down when all the material content of the universe is piled into a single point. 
Whenever the particle is actually a planet or a star, it has material incompressibility, which eventually amounts to quantum degeneracy pressure. 
While this effect can be overcome, a generally-relativistic regime -- black hole formation -- takes over, so classical Newtonian point-particle models 
cease to be appropriate, so there is a physical reason to excise or unfold.  
While these consideration apply to any point-or-particle coincidence-or-collision -- maximal or partial -- 
some further Background Independence considerations apply solely to the maximal case, since here in which all the material content of the universe is piled up.
This gives further reasons for specifically maximal coincidence-or-collisions to be removed.\footnote{Mathematically, the $N = 2$ coincidence-or-collision moreover 
%OOOOOOOOOOOOOOOOOOOOOOOOOOOOOOOOOOOOOOOOOOOOOOOOOOOOOOOOOOOOOOOOOOOOOOOOOOOOOOOOOOOOOOOOOOOOOOOOOOOOOOOOOOOOOOOOOOOOOOOOOOOOOOOOOOOOOOOOOOOOOOOOOOOOOOOOOOOOOOOOOO
behaves like subsequent {\sl partial} rather than maximal such.
Thus the full list of reasons to excise maximal coincidence-or-collisions only applies from the $d = 2$, $N = 3$ `triangle' configurations upwards.}
%OOOOOOOOOOOOOOOOOOOOOOOOOOOOOOOOOOOOOOOOOOOOOOOOOOOOOOOOOOOOOOOOOOOOOOOOOOOOOOOOOOOOOOOOOOOOOOOOOOOOOOOOOOOOOOOOOOOOOOOOOOOOOOOOOOOOOOOOOOOOOOOOOOOOOOOOOOOOOOOOOO

\m 

\n{\bf Remark 9} In $d = 3$, some works furthermore excise the collinear configurations. 
For $N = 3$, these amount to a further stratum: the planar edge of the hemisphere of mirror image identified triangular shapes in the pure-shape case, 
or the corresponding plane (with O point deleted as being yet another separate stratum) in the scaled case. 
Now there is {\sl nothing} physically wrong with those collinear configurations that are not concurrently partial collisions, 
so excision or unfolding in this case lacks physical justification.
%
% We note that these examples are moreover prototypical for further significant such in the wider range of Mechanics Theories alluded to above.  
 
\m

\n{\bf Remark 10} All in all, maximal coincidence-or-collisions, for $d \geq 2$ and $N \geq 3$ in particular, have various mathematical pathologies associated with them,  
but can be conveniently argued away by excision or unfolding, which moreover have at least some qualitative physical justification.  
For a few decades, this (and more questionable uses of these same approaches on collinear configurations) sufficed.  
The next two examples below, however, show that current conceptions of Excise Strata and Unfold Strata approaches 
are not well-suited to handle all the stratification that can occur in the somewhat larger arena of relational spaces.
These further examples -- Type 3 -- are moreover naturally-ocurring and well-motivated shape spaces, namely those of Image Analysis and Computer Vision.

%===================================================================================================================================================================
\subsection{Loss of Hausdorffness}
%===================================================================================================================================================================

In the current Article, we focus in particular on whether relational quotients are Hausdorff-separable; 
this requires first contemplating what form weaker notions of separation take.  

\m 

\n{\bf Definition 2} A topological space $\FrX$ is {\it Fr\'{e}chet} \cite{Willard, Engelking} alias {\it accessible} 
if whichever 2 distinct points $x \neq y$ in $\FrX$ are separated by there being 
\be 
\mbox{at least one open set \m $\FrU$ \m such that \m $x \in \FrU$ \m but \m $y \not{\hspace{-0.05in}\in} \m \FrU$} \m . 
\ee 
\n{\bf Remark 11} This is to be contrast with Hausdorffness, for which {\sl two distinct} open sets are involved in a symmetric manner. 
Uniqueness for limits of sequences need no longer hold, but all points are closed sets.

\m

\n{\bf Definition 3} A topological space $\FrX$ is {\it Kolmogorov} \cite{Willard, Engelking} 
if whichever 2 distinct points $x \neq y$ in $\FrX$ are {\it topologically distinguishable}, meaning that  
\be 
\mbox{$\exists$ \m \m at least one open set \m $\FrU$ \m  such that \m $x \in \FrU$ \m but \m $y \not{\hspace{-0.05in}\in} \m \FrU$}  \mma  
                                                 \mbox{or } \m  \mbox{$x \in \FrU$ \m  but \m $y \not{\hspace{-0.05in}\in} \m \FrU$}  \m .
\ee 
\n{\bf Remark 12} Now furthermore points need no longer be closed sets; 
the Sierpi\'{n}ski topology on two points, with one closed and one not, is a well-known example and the smallest merely-Kolmogorov topology.  
Kolmogorov separability is moreover much weaker than, and {\sl qualitatively distinct} from, the further sequence of types of separability beginning with with Fr\'{e}chet and Hausdorff. 

\m 

\n{\bf Remark 13} There is moreover an identification construct \cite{Willard} by which Kolmogorovness is a {'guaranteed' minimum standard} of separability. 
[Some modelling situations might prohibit making this identification.]

%===================================================================================================================================================================
\subsection{Type 3: stratified manifolds with widespread obstructions to Analysis}
%===================================================================================================================================================================

\n{\bf Example 4} For nontrivial Affine Shape Theory, $d \geq 2$ is required for nontriviality.  
For $d = 2$ itself, $N = 4$ is then minimal to have a nontrivial theory: the affine quadrilaterals in the plane \cite{GT09, Aff}.
In this model, David Groisser and Hemant Tagare \cite{GT09} have shown that each of collinear shapes C and non-collinear shapes G form their own $\mathbb{RP}^2$ stratum, 
\be 
\FrS(2, 4; Aff(2))  \es  \mathbb{RP}^2 \m \disjoint \m \mathbb{RP}^2           \m , 
\ee 
with every point in C moreover lying arbitrarily close to every point in G.
This renders the affine shape space merely Kolmogorov-separated.  

\m 

\n Affine shape spaces having multiple strata rests on $GL(d, \mathbb{R})$ possessing a $SO(d)$ subgroup, a feature which clearly persists with increasing $d$ and $N$.  
The merely-Kolmogorov separability itself also persists \cite{GT09} in these larger models.

\m 

\n The affine quadrilateral already illustrates non-principal strata being equidimensional with principal strata.  
In some of the larger models, moreover, nonprincipal strata can even exceed the dimension of the corresponding principal strata.  

\m 

\n{\bf Example 5} Merely-Kolmogorov separation in fact already occurs in the version of Kendall's Shape Theory in which one 
insists in keeping the maximal coincidence-or-collision O amongst the shapes. 
For in this case, O is arbitrarily close to every point in the phase space. 
This is however another instance of isolated bad point (Type 2) and usually simply treated by well-motivated excision. 
However, neither this classification nor this solution transcend however to our affine example above (or our projective example below).

\m 

\n{\bf Example 6} Projective Geometry is well-known to cure Affine Geometry of various imperfections \cite{HC32, Hartshorne}.
However, passing to Projective Shape Theory {\sl does not} alleviate the corresponding Shape Theory's shape space's merely-Kolmogorov separation.
This was recently shown by Florian Kelma, John Kent and Thomas Hotz \cite{KKH16}, 
who moreover consider an accessible subspace formed by including only a more limited selection of projective shapes.  

\m

\n{\bf Remark 14} Substantial technical problems are thus posed by these affine and projective examples (see Remark 15 below for more details). 
In view of this, we formulate a second, more specific and severe, global problem as follows.  

\m
 
\n{\bf Global Problem 2} Configuration space reductions can kick one out of the class of manifolds into the class of not Hausdorff but 
merely Kolmogorov-separated stratified manifolds. 
In the current study of Relational Theories, this means that the relational spaces -- shape(-and-scale) spaces -- are capable of being spaces of this kind. 

\m 

\n{\bf Remark 15} The current Article moreover explores how Kendall's Shape Theory itself manages to be immune from this phenomenon. 
The resulting topological spaces level considerations moreover point to Hausdorff separation seldom being guaranteed in the general study of relational spaces.  

\m 

\n{\bf Remark 16} The excise and unfold strategies of Sec 4 work for stratified manifolds that are `analytically nice', whether or not modulo isolated `bad points' 
as exemplified by the maximal coincidence-or-collision.   
There are however  further problems with these strategies in the wider context of the current subsection. 

\m 

\n 1) That nonprincipal strata can be of {\sl equal or higher dimension} to the principal stratum presents a further problem for Excise Strata.  
Excise Strata is a strategy which pre-dates observation of nonprincipal strata that are equal- or higher-dimensional relative to the corresponding principal strata.

\m 

\n 2) One way to Unfold Strata for Affine Shape Theory is to send it back to Kendall's similarity model.  
In 2-$d$, this is a complete unfolding, returning a $\mathbb{CP}^{N - 2}$ shape space manifold, whereas in 3-$d$ it is a partial unfolding: 
$\FrS(d, N)$ still has strata, albeit now of a nicer LCHS type.  
This unfolding however fails to apply Image Analysis' well-motivated modelling assumptions.  
Unfold Strata by itself may also fails to address the issue of separability between strata; 
it is a strategy which pre-dates observation of merely-Kolmogorov separation between strata. 

\m 

\n 3) This leaves us with Accept Strata, albeit this now amounts to accepting merely-Kolmogorov separation can occur between strata. 
This lies well beyond the remit of currently-available calculational tools \cite{Pflaum, Kreck, SniBook} put forward for handling stratified manifolds.

%===================================================================================================================================================================
%===================================================================================================================================================================
\section{Fibre bundles cease to suffice upon working over stratified manifolds}
%===================================================================================================================================================================
%===================================================================================================================================================================

{\bf Structure 1} {\it Fibre bundles} \cite{IshamBook2, Husemoller} extend study of a given manifold by attaching identical fibres to each of its points, 
forming a total space.
Points on each fibre moreover project down onto each of the original (`base') manifolds' points.
Some global properties of manifolds follow from considerations of sections: cuts through the total space to which each fibre contributes just one point.  

\m 

\n When one is dealing instead with a stratified manifold, fibre bundles \cite{IshamBook2, Husemoller} in general cease to suffice because desirable attached objects 
are now heterogeneous stratum by stratum.  
For instance, charts and tangent spaces in general differ in dimension from stratum to stratum.  

\m 

\n{\bf Structure 2} {\it General bundles} \cite{IshamBook2, Husemoller} are minimal as regards handling heterogeneous attached objects. 

\m

\n{\bf Structure 3} {\it Presheaves} \cite{MM92, Sheaves} can also handle heterogeneous attached objects; 
these amount to a mathematical reconceptualization in which restriction maps play a central role. 

\m 

\n{\bf Structure 4} Sheaves are then a furtherly structured approach possessing two further useful properties a `local to global' gluing and a `global to local' condition.  
{\sl Sheaves are tools for tracking locally defined entities by attaching them to open sets within a topological space.}
Attaching locally defined entities to open sets within a stratified manifold is, straightforwardly, a subcase or application of this. 
We need not further detail sheaves for the purposes of the current Article; see \cite{Wells, Ghrist, ABook, Wedhorn} for outlines 
and \cite{Sheaves, Hartshorne} for detailed texts.

\m 

\n{\bf Structure 5} Differential spaces \cite{Sikorski, LS08, SniBook} are the structure actually evoked in the current Article, 
with Sec 8.7 detailing their definition.  

\m 

\n Using sheaves versus differential spaces is a major difference between Kreck and \'{S}niatycki's programs for handling stratified manifolds. 
Pflaum and Kreck, moreover, both consider 
\be 
\mbox{(stratified manifold, sheaf)} 
\ee
pairs for their slightly different range of stratified spaces (LCHP versus LCHS).  
Kreck coined the word {\it stratifold} for his version of such pairs \cite{Kreck}.

%===================================================================================================================================================================
%==================================================================================================================================================================
\section{Stratification versus (Killing-based) genericity}
%==================================================================================================================================================================
%===================================================================================================================================================================

\n Before proceeding with the main results of the current Article, 
let us note that an alternative to stratification is incorporating small defect \cite{A-Generic} into one's manifold model of carrier space. 
This leaves one's carrier space with no generalized symmetries, by which the only relational space is the trivial constellation space.  

\m 

\n This can be viewed as a type of Unfold Strata, 
and indeed one which can involve more realistic modelling, since small defects abound in actual systems being modelled. 
This approach also has a rather longer-serving GR configuration space \cite{DeWitt67, DeWitt70} parallel \cite{FM96}; see \cite{A-Generic} and Article III for further discussion and comparison.  

\m

\n In both cases, moreover, models lacking symmetry are generic (see Articles I and III respectively).
So nontrivial Kendall-type Relational Theories are of measure zero over the space of all possible absolute space manifolds, 
and nontrivial GR configuration spaces are of measure zero over the space of all possible topological 3-manifolds.  

\m  

\n Ascribing to this section's incorporating small defects approach is a possible response to models which, 
without small defects, give merely-Kolmogorov separated stratified manifolds.  
The space and configuration space thus built 
can however no longer be studied by the exact symmetry methods that Theoretical Physics (and to some extent Applied Mathematics) has hitherto greatly relied upon.  
This is moreover the situation one has to face once one is working with a {\sl generic} carrier space.
There is moreover a further problem in the `incorporating small defects' approach. 
I.e.\ the requirement to show that, if the small defects one includes are arbitrary, the resulting Relational Theory can make at least approximate 
predictions independent of the form of the defects incorporated. 
Even if the defects one includes are observed, one would need to show stability to the inclusion of even smaller defects, which, 
in certain locations and/or beneath certain scales, would become arbitrary.  

\m 

\n For the rest of the current Article, we concentrate instead the inner workings of the case of carrier spaces with generalized symmetries that lead to strata.

%===================================================================================================================================================================================
%===================================================================================================================================================================================
\section{Compactness selection principles}
%===================================================================================================================================================================================
%===================================================================================================================================================================================

%===================================================================================================================================================================
\subsection{Kendall's Shape Theory}
%===================================================================================================================================================================

\n{\bf Remark 1} We first observe two instances of compactness in Kendall's Shape Theory.  

\m 

\n Firstly, the similarity group 
\be  
Sim(d)  \es Tr(d) \rtimes (Rot(d) \times Dil) 
        \es \mathbb{R}^d \rtimes (SO(d) \times  \mathbb{R}_+)
\ee  
itself is noncompact. 
However, 

\m 

\n 1) the centre of mass frame map removes $Tr(d)$.    

\m    

\n 2) The unit sphere          map removes $Dil$.  

\m  

\n Thus all of $Sim(d)$'s noncompact generators can be removed.  

\m 

\n{\bf Remark 2} This leaves us quotienting a compact Lie group from a compact manifold,
\be 
\frac{\FrM_{\sC}}{\lFrg_{\sC}}  \m , 
\ee 
where the C subscript stands for `compact'.

\m 

\n{\bf Remark 3} More specifically, this leaves us quotienting the compact rotation group from the compact preshape sphere:
\be 
\FrS(d, N)  \:=  \frac{\FrP(d, N)}{Rot(d)}
            \es  \frac{\mathbb{S}^{n \, d - 1}}{SO(d)}  \m .  
\ee 
for $n := N - 1$.

\m 

\n On the one hand, for $d = 1$, 2, these are manifolds: $\mathbb{S}^{n - 1}$ and, by the generalized Hopf map, $\mathbb{CP}^{n - 1}$.  

\m 

\n On the other hand, for $d \geq 3$, strata appear, but the relational space remains LCHS.

%===================================================================================================================================================================
\subsection{Maximal compact subgroups}\label{MCG}
%===================================================================================================================================================================

\n{\bf Remark 4} Extending to Affine Shape Theory, we still have the compact preshape space, but the group being quotiented out, 
after 1) and 2), reduces to $SL(d, \mathbb{R})$, which is noncompact.  
 
\m 

\n{\bf Remark 5} Conformal and Projective Shape Theories' algebraic structure is such that piecemeal removal of $Tr(d)$ or $Dil$ is not possible \cite{A-Brackets}. 
Thus we have the form 
\be 
\frac{\FrM_{\sN}}{\lFrg_{\sN}}  \m , 
\ee 
where the N subscript stands for `noncompact'.
\n $SO(d)$ is the {\it maximal compact subgoup} of $SL(d, \mathbb{R})$, $Proj(d)$, and $Conf(d)$, which are moreover related by the partial order in Fig \ref{Cpct1}.a). 
See Fig \ref{Cpct1}.b) for the corresponding 1-$d$ subcase.  

\m 

\n The 2-$d$ case involves a finitely generated subgroup of $Conf(2)$ rather than the whole of $Conf(2)$; 
the full $Conf(2)$, moreover, does not support a Shape Theory.

%===================================================================================================================================================================
\subsection{Compactness is productive and quotientive}
%===================================================================================================================================================================

\n We next turn to two basic properties of compactness, which help in further determining which numerator-manifolds and quotients are compact. 

\m 

\n {\bf Proposition 1} Compactness is productive \cite{Lee1, Munkres, Willard}. 

\m 

\n{\bf Remark 6} In the (Scale-and-)Shape Theory setting, moreover, 
we only need concern ourselves with the simpler case of finite products (rather than infinite ones).

\m

\n{\bf Corollary 5} By (\ref{Constell}), compact carrier space $\FrM_{\sC}$ thus implies compact constellation space,  
\be 
\FrQ_{\sC}(d, N) = \bigtimes_{I = 1}^N \FrM_{\sC}  \m . 
\ee  
\n Thus, among the current Article's examples, circle, torus, sphere and real-projective carrier spaces beget compact constellation spaces. 

\m 

\n{\bf Corollary 6} The circle $\mathbb{S}^1$ and torus $\mathbb{T}^d$ relative spaces are compact. 

\m 

\n{\bf Remark 7} On the one hand, $\mathbb{S}^1$, $\mathbb{T}^d$, $\mathbb{S}^d$, and $\mathbb{RP}^d$ join 
'$\mathbb{R}^d$ with up to affine transformations quotiented out' 
in satisfying the condition of their relational spaces being formulable in terms of a compact numerator-manifold.
This covers Examples 0-3 and 6-9 of Article I.
\m 

\n On the other hand, hyperbolic space $\mathbb{H}^d$ and Minkowski spacetime $\mathbb{M}^D$ (Examples 10 and 12 of Article I) are non-compact, 
with the first not admitting a similarity Killing vector and the second not forming a compact preshape space due to indefiniteness.  

\m 

\n {\bf Proposition 2} Compactness is quotientive \cite{Lee1}.

\m  

\n{\bf Remark 8} This is a way in which compact numerator-manifolds protect ensuing relational spaces.

\m 

\n In particular, Kendall's preshape sphere $\mathbb{S}^{n \, d - 1}$ is compact and thus protects subsequent quotients.\footnote{Connectedness and path-connectedness are moreover 
%OOOOOOOOOOOOOOOOOOOOOOOOOOOOOOOOOOOOOOOOOOOOOOOOOOOOOOOOOOOOOOOOOOOOOOOOOOOOOOOOOOOOOOOOOOOOOOOOOOOOOOOOOOOOOOOOOOOOOOOOOOOOOOOOOOOOOOOOOOOOOOOOOOOOOOOOOOOOOOOOOOO
also all of productive, quotientive and properties of the $k$-sphere \cite{Lee1}.  
This gives us a further simple qualitative grasp over ensuing relational spaces.
Connectedness is also an enabler of Theorems of Analysis: the Intermediate Value Theorem respectively.

\m 

\n The exceptional geometrical tractability of spheres is a final source of tractability arising from Kendall's preshape sphere \cite{Kendall, FileR}, 
in particular in 2-$d$ by admitting the generalized Hopf map to $\mathbb{CP}^{N - 2}$.}
%OOOOOOOOOOOOOOOOOOOOOOOOOOOOOOOOOOOOOOOOOOOOOOOOOOOOOOOOOOOOOOOOOOOOOOOOOOOOOOOOOOOOOOOOOOOOOOOOOOOOOOOOOOOOOOOOOOOOOOOOOOOOOOOOOOOOOOOOOOOOOOOOOOOOOOOOOOOOOOOOOOO 
%
This protection holds for quotients by $Sim(d)$ and $Aff(d)$, and for further discrete quotients in each case.  

\m 

\n $Conf(d)$ and $Proj(d)$ (Examples 4 and 5 of Article I) however have no corresponding compact numerator-manifold, 
by not being able to separate out a preshape sphere, by Remark 5.

%===================================================================================================================================================================
\subsection{Quotienting by proper group actions}
%===================================================================================================================================================================

\n We proceed further via first pointing out a result that involves the right classes of numerator-entity -- a smooth manifold -- 
                                                                                  and of denominator-entity: a Lie group, 
																				  but which is nonetheless too specialized for the current application.  

\m 

\n{\bf Definition 1} An action of a group $\lFrg$ on a space $X$ is a map 
\be 
\alpha \m : \m \m X \times \lFrg \longrightarrow X 
\ee 
such that 
\be 
x \cdot e = x   \m \forall x \in M \m \m \mbox{(identity)} \m , 
\ee 
and 
\be 
x \cdot (g_1 \, g_2) = (x \cdot g_1) \cdot g_2  \m \forall x \in M \mma \forall g_1 \m , \m g_2 \in \lFrg \m \mbox{ (associativity)}  \m . 
\ee 
\n{\bf Theorem 1 (Quotient Manifold Theorem)} \cite{Lee2} Let $\FrM$ be a smooth manifold and $\lFrg$ be a Lie group acting thereupon, according to 
\be 
\Phi: \lFrg \times \FrM \m \longrightarrow \m \FrM   \m . 
\label{Proper}
\ee  
Then if $\lFrg$'s action is smooth, proper and free, 
\be 
\mbox{the corresponding quotient space } \m \frac{\FrM}{\lFrg} \m \mbox{ is itself a manifold} \m .  
\ee 
\n This is too specialized since our interest is in determining which stratified manifolds arising from quotienting -- a more general outcome -- are Hausdorff. 
The Quotient Manifold Theorem is however a criterion {\sl guaranteeing} manifoldness, and thus in particular Hausdorffness.  

\m 

\n{\bf Remark 9} We are thus to progress by weakening the Quotient Manifold Theorem's premises.  
Let us first examine the definitions of `proper' and `free' actions \cite{Lee1}. 

\m 

\n{\bf Definition 2} A group action of $\lFrg$ on a topological space $\FrX$ is {\it proper} if each compact subset's 
\be 
\FrK \subseteq \FrX
\ee  
action inverse-image 
\be
\Phi^{-1}(\FrK)  
\ee 
is itself compact \cite{Lee1}.  

\m 

\n{\bf Definition 3} A group action is {\it free} if the sole element of $\lFrg$ fixing any point in $\FrX$ is the identity element.    

\m 

\n{\bf Remark 10} On the one hand, we identify freeness as an unwanted stringency as regards the study of stratified manifolds. 

\m 

\n{\bf Remark 11} On the other hand, we identify properness as a derivative notion of compactness -- more precisely a relative notion of compactness \cite{Wedhorn} -- 
and our main idea is to {\sl keep} this feature.  
We do so moreover in the context of the following further result. 

\m  
   
\n{\bf Lemma 1} \cite{Lee1} Every continuous action of a compact topological group on a Hausdorff space is proper. 
%
% p 319

\m 

\n{\bf Corollary 7} Compact geometrical automorphism groups act properly on constellation spaces.  

\m 

\n{\u{Proof}} On the one hand, constellation spaces are manifolds and in particular are thus Hausdorff spaces. 

\m 

\n On the other hand, automorphism groups are Lie groups and thus topological groups. $\Box$

%==================================================================================================================================================================
\subsection{Protective theorems in the case of quotienting out by compact Lie groups }
%==================================================================================================================================================================

\n{\bf Remark 12} Closed map techniques \cite{Lee1, Willard} applied to quotient maps yield the following results. 

\m 

\n{\bf Theorem 2} \cite{Lee1} For $\FrX$ HLC and $\lFrg$ a topological group acting continuously and properly, the corresponding 
\be 
\mbox{quotient space} \m \m \frac{\FrX}{\lFrg} \m \m  \mbox{is Hausdorff} \m .
\ee 
%
% p 320. 

\m 

\n{\bf Corollary 8} For $\FrM$ a manifold and $\lFrg$ a compact topological group acting continuously, the corresponding 
\be
\mbox{quotient space} \m \m \frac{\FrM}{\lFrg} \m \m  \mbox{is Hausdorff} \m . 
\ee 

\m 

\n{\u{Proof}} Manifolds are both Hausdorff and locally-compact. 

\m 

\n Lemma 1 moreover applies, ensuring that the action in question is proper.

\m 

\n Theorem 2 then applies. $\Box$ 

\m 

\n{\bf Corollary 9} Relational spaces formed by continuous action of a compact Lie group of geometrical automorphisms are Hausdorff. 

\m 

\n{\u{Proof}} The numerator-object is constellation space, and thus a manifold. 
Then apply Corollary 8. $\Box$

\m 

\n{\bf Remark 13} We have moreover the following more powerful result.  

\m 

\n{\bf Theorem 3} (Exercise 4.8 on p.\ 199 of \cite{Munkres}) Quotienting a topological space $\FrX$ 
by the continuous action of a compact topological group $\lFrg$ preseves Hausdorffness, second-countability and local compactness. 

\m 

\n{\bf Corollary 10} For $\FrM$ a manifold and $\lFrg$ a compact Lie group, the corresponding 
\be 
\mbox{quotient space} \m \m \frac{\FrM}{\lFrg} \m \m \mbox{is LCHS} \m .
\ee 
\m 

\n\n{\u{Proof}} Manifolds are LCHS spaces and Lie groups are topological groups, so Theorem 2 applies. $\Box$ 

\m 

\n{\bf Corollary 11} Relational spaces formed by action of a compact Lie group of geometrical automorphisms are LCHS.  

\m 

\n\n{\u{Proof}} The numerator-object is constellation space, and thus a manifold, so Corollary 10 applies. $\Box$ 

\m 

\n{\bf Remark 14} This permits us to establish the whole LCHS package of analytic nicety in the case of quotienting by a compact Lie group, 
thus constitutes a protective theorem for this case.

\m 

\n{\bf Theorem 4}  If an LCHS topological group $\lFrg$ acting properly on LCHS space $X$, then 
\be 
\frac{X}{\lFrg} \m \mbox{ is itself LCHS} \m .  
\ee 
%
%%%%%%%%%%%%%%%%%%%%%%%%%%%%%%%%%%%%%%%%%%%%%%%%%%%%%%%%%%%%%%%%%%%%%%%%%%%%%%%%%%%%%%%%%%%%%%%%%%%%%%%%%%%%%%%%%%%%%%%%%%%%%%%%%%%%%%%%%%%%%%%%%%%%%%%%%%%%%%%%%%%%%%%%%%%%%%%%%%%%
% \n + Proper action $\Rightarrow$ $\exists$ slices $\Rightarrow$ one can partially-order the orbit types \cite{Adams}.  
%
% D.V. Alkseevsky, ``On a Proper Action of a Lie Group" USp. Mat. Nauk. {\bf 34} 219 (1979).
%
% An action is proper iff there is a $\lFrG-invariant Riemannian metric on $X$.  
%
% In the non-proper case, there is no slice in general.  
%
% Palais, Am. Math. {\bf 73} 295 (1961): there exist slices for actions of noncompact Lie groups.
%%%%%%%%%%%%%%%%%%%%%%%%%%%%%%%%%%%%%%%%%%%%%%%%%%%%%%%%%%%%%%%%%%%%%%%%%%%%%%%%%%%%%%%%%%%%%%%%%%%%%%%%%%%%%%%%%%%%%%%%%%%%%%%%%%%%%%%%%%%%%%%%%%%%%%%%%%%%%%%%%%%%%%%%%%%%%%%%%%%%
%
\n{\bf Remark 15} LCHS spaces and LCHS topological groups generalize manifolds and Lie groups respectively, so this is more general than Theorem 3.  

\m 

\n{\bf Remark 16} Adding discrete transformations to a continuous Lie group can breach connectedness -- giving multiple connected components -- 
but not compactness, by which the results so far in this section carry over to this case.  
Our next result, on the nature of each stratum, moreover explicitly evokes a connected Lie group.  

\m 

\n{\bf Theorem 4} \cite{LS08, DK00} Given a proper action (\ref{Proper}) of a connected Lie group $\lFrg$ on a manifold $\FrM$, the corresponding 
\be 
\mbox{quotient } \m  \frac{\FrM}{\lFrg} \m  \mbox{ is stratified by orbit type} \m . 
\ee 
%
% p 13 of LS08, not stated as a Theorem since it was already known.  
%
\n{\bf Remark 17} By Lemma 1, we can moreover rearrange this into the following form. 

\m 

\n{\bf Corollary 12} Let $\lFrg$ be a compact connected Lie group acting smoothly on a manifold $\FrM$. 
Then  
\be 
\frac{\FrM}{\lFrg} \m \mbox{ is stratified by orbit type} \m . 
\ee

%===================================================================================================================================================================================
\subsection{Examples of reducible noncompactness}
%===================================================================================================================================================================================

\n{\bf Example 1} For the translational group, itself noncompact, $\Phi^{-1}$ sends intervals to translated intervals -- which are still compact -- and so is a proper action. 

\m

\n \cite{GT09} does not state nonproper action to imply non-Hausdorff quotient. 
This is indeed a possible -- and perhaps common -- phenomenon, but does not always occur. 
%
%%%%%%%%%%%%%%%%%%%%%%%%%%%%%%%%%%%%%%%%%%%%%%%%%%%%%%%%%%%%%%%%%%%%%%%%%%%%%%%%%%%%%%%%%%%%%%%%%%%%%%%%%%%%%%%%%%%%%%%%%%%%%%%%%%%%%%%%%%%%%%%%%%%%%%%%%%%%%%%%%%%%%%%%%%%%%%%%%%%%
% For take $\mathbb{R}^{Nd}/Tr(d$. 
%
% $Tr(d)$ is not compact but its action is still proper. 
%%%%%%%%%%%%%%%%%%%%%%%%%%%%%%%%%%%%%%%%%%%%%%%%%%%%%%%%%%%%%%%%%%%%%%%%%%%%%%%%%%%%%%%%%%%%%%%%%%%%%%%%%%%%%%%%%%%%%%%%%%%%%%%%%%%%%%%%%%%%%%%%%%%%%%%%%%%%%%%%%%%%%%%%%%%%%%%%%%%% 
%
For working with normalizable similarity shapes gives a Hausdorff quotient.  
 
\m  

\n{\bf Example 2} For the dilational group, $\Phi^{-1}$ sends $\mathbb{S}^{n d - 1}$ to $\mathbb{R}^{nd}$, noncompact.  
So this action is not proper. 
This case does give also that the maximal collision is arbitrarily close to all other shapes. 
But this is a single point, and its excision leaves one with the sphere and is moreover justifiable on physical grounds.  
With this excision done, dilations acting on Euclidean space exemplify reducible noncompactnesses, since the corresponding quotient is rerepresentable in the form 
\be 
\frac{\FrM_{\sC}}{\lFrg_{\sC}}  \m .
\ee  
% + Centralize introduction of this notation.

\m 

\n $X/\lFrg$ acting nonproperly can give Hausdorff by cancellation examples, which serve as counterexamples to these being a general theorem.  
%
% + Centralize
%
%%%%%%%%%%%%%%%%%%%%%%%%%%%%%%%%%%%%%%%%%%%%%%%%%%%%%%%%%%%%%%%%%%%%%%%%%%%%%%%%%%%%%%%%%%%%%%%%%%%%%%%%%%%%%%%%%%%%%%%%%%%%%%%%%%%%%%%%%%%%%%%%%%%%%%%%%%%%%%%%%%%%%%%%%%%%%%%%%%%%
% \n + p2 Schmitt online: continuous Lie group action means that $G/X$ is Hausdorff.
%%%%%%%%%%%%%%%%%%%%%%%%%%%%%%%%%%%%%%%%%%%%%%%%%%%%%%%%%%%%%%%%%%%%%%%%%%%%%%%%%%%%%%%%%%%%%%%%%%%%%%%%%%%%%%%%%%%%%%%%%%%%%%%%%%%%%%%%%%%%%%%%%%%%%%%%%%%%%%%%%%%%%%%%%%%%%%%%%%%%

%==================================================================================================================================================================
\subsection{Subcartesian differential spaces} 
%==================================================================================================================================================================

\n Having outlined applying Kreck's approach to LCHS stratified manifolds to Shape(-and-Scale) Theory in \cite{ABook, A-Generic},  
on the present occasion we elect instead to outline part of \'{S}niatycki's approach \cite{LS08, SniBook}.  
This permits a somewhat more detailed description of the nature of each stratum.  

\m 

\n{\bf Definition 4} A {\it differential space} is a topological space $\FrX$ 
equipped with the differential structure of a subring of continuous functions satisfying the following axioms \cite{LS08}.

\m 

\n{\bf Differential Spaces 1} The family of sets 
\be 
\left\{ \, f^{-1}((a, b)) \, | \, f \in {\cal C}^{\infty}(\FrX), a, b \in \mathbb{R} \, \right\} 
\ee
is a sub-base for $\FrX$'s topology. 

\m

\n\n{\bf Differential Spaces 2} Suppose a function $f$ on $\FrX$ is such that 
\be 
\forall \m x \in \FrX \m \exists \m \mbox{ (open neighbourhood of $x$ \, : \m $\FrU_x \subseteq Q$) }
\ee  
and a function 
\be 
f_x \in {\cal C}^{\infty}(\FrX)
\ee 
such that the alignment of restrictions\footnote{Both differential spaces and presheaves are thus rooted in restriction maps.} 
\be 
f|_{\FrU_x}  \es  f_x|_{\FrU_x}                          \m ,
\ee
holds. 
Then 
\be 
f \, \in \, {\cal C}^{\infty}(\FrX) \m \mbox{ (closure)}   \m .  
\ee 
\n\n{\bf Differential Spaces 3}
\be 
\mbox{$\forall \m n \m \in \m \mathbb{N}$ \mma $f_i \, \in \, {\cal C}^{\infty}(\FrX)$ \m ($i = 1$ to $n$) \m \m and \m \m $F \in {\cal C}^{\infty}(\mathbb{R}^k)$} \m , 
\ee 
\be 
F(f_1, \, . \m . \m . \m  , \, f_n) \m \in \m {\cal C}^{\infty}(\FrX) \m \mbox{ (closure under composition)}   \m . 
\ee
\n{\bf Remark 18} These spaces have the advantage of including some stratified spaces while also allowing for a notion of derivation to exist: 
a partial weakening of manifolds' tangent bundle. 

\m 

\n Noting moreover that differential spaces' definition makes no use of our target good behaviour -- Hausdorffness -- 
we strengthen our requirements with the following adjective.

\m 

\n{\bf Definition 5} A {\it subcartesian space} is a Hausdorff differential space $\FrX$ such that 
each point $x \in \FrX$ possesses a neigbourhood diffeomorphic to a subset of a Cartesian space $\mathbb{R}^K$.  

\m 

\n{\bf Remark 19} The general idea then is that a subcartesian space is suitable for much Analysis, including notions of Calculus and of Differential Topology.
Such are rather well suited to Physics (and continuum Probability and Statistics).
Subcartesian spaces are moreover relevant to the current study via the following result (which we again slightly rearrange using Lemma 1). 

\m

\n{\bf Theorem 6} \cite{LS08} Given a proper action (\ref{Proper}) of a connected Lie group $\lFrg$ on a manifold $\FrM$, the corresponding 
\be
\mbox{quotient space} \m \frac{\FrM}{\lFrg} \m \mbox{ is a subcartesian space} \m .
\ee
\n{\bf Corollary 13} Let $\lFrg$ be a compact connected Lie group acting smoothly on a manifold $\FrM$. 
Then the corresponding
\be
\mbox{quotient space} \m \frac{\FrM}{\lFrg} \m \mbox{ is a subcartesian space} \m .
\ee
\n{\bf Remark 20} The promised enhanced result about the nature of each stratum is then as follows, including once again rearrangement by Lemma 1. 

\m

\n{\bf Theorem 7} \cite{LS08} For a proper action of a connected Lie group $\lFrg$ on a manifold $\FrM$, the corresponding 
\be 
\mbox{quotient } \m \frac{\FrM}{\lFrg} \m \mbox{ 's stratification by orbit type coincides with the partition of } \m \FrM  
\ee 
\be 
\mbox{by the family of orbits of all vector fields on } \m  \frac{\FrM}{\lFrg} \m .  
\ee 
\n{\bf Corollary 14} Let $\lFrg$ be a compact connected Lie group acting smoothly on a manifold $\FrM$. 
\be 
\mbox{Then the corresponding quotient } \m \frac{\FrM}{\lFrg} \m \mbox{ 's stratification by orbit type coincides with the partition of } \m \FrM  
\ee 
\be 
\mbox{by the family of orbits of all vector fields on } \m  \frac{\FrM}{\lFrg} \m .  
\ee
\n{\bf Remark 21} Relational spaces that can be phrased in terms of a compact Lie group $\lFrg$ are thus guaranteed not only analytic protection 
but also a substantial degree of differential-geometric familiarity.

%===================================================================================================================================================================
\subsection{Classification of Article I's wider range of relational theories by compactness criteria} 
%===================================================================================================================================================================

This classification is given in Fig 1. 

\m 

\n{\bf Remark 22} We cannot discount possibility of further cancellations in noncompact/noncompact presentations giving 'reducible noncompactness': 
reducible by rerepresentation of the quotient.   
We talk however about the situation of quotienting out by noncompact groups in the absence of such resolution. 
I.e.\ {\it nonresolvably noncompact} geometrical automorphism groups being quotiented out.   
On the other hand, compact automorphism groups and resolvably-noncompact groups give the simplest Type i) strata.
`Resolvable' here refers to passing to the centre of mass frame nullifying the translations, or their compactified equivalents in $\mathbb{T}^d$. 
The aim is then to either find a counterexample or to rest this observation on rigorous theorems.  
%
%FFFFFFFFFFFFFFFFFFFFFFFFFFFFFFFFFFFFFFFFFFFFFFFFFFFFFFFFFFFFFFFFFFFFFFFFFFFFFFFFFFFFFFFFFFFFFFFFFFFFFFFFFFFFFFFFFFFFFFFFFFFFFFFFFFFFFFFFFFFFFFFFFFFFFFFFFFFFFFFFFFFFFFFFFFFFFFFFFFFFFFFFF
{\begin{figure}[ht]
\centering
\includegraphics[width=1.0\textwidth]{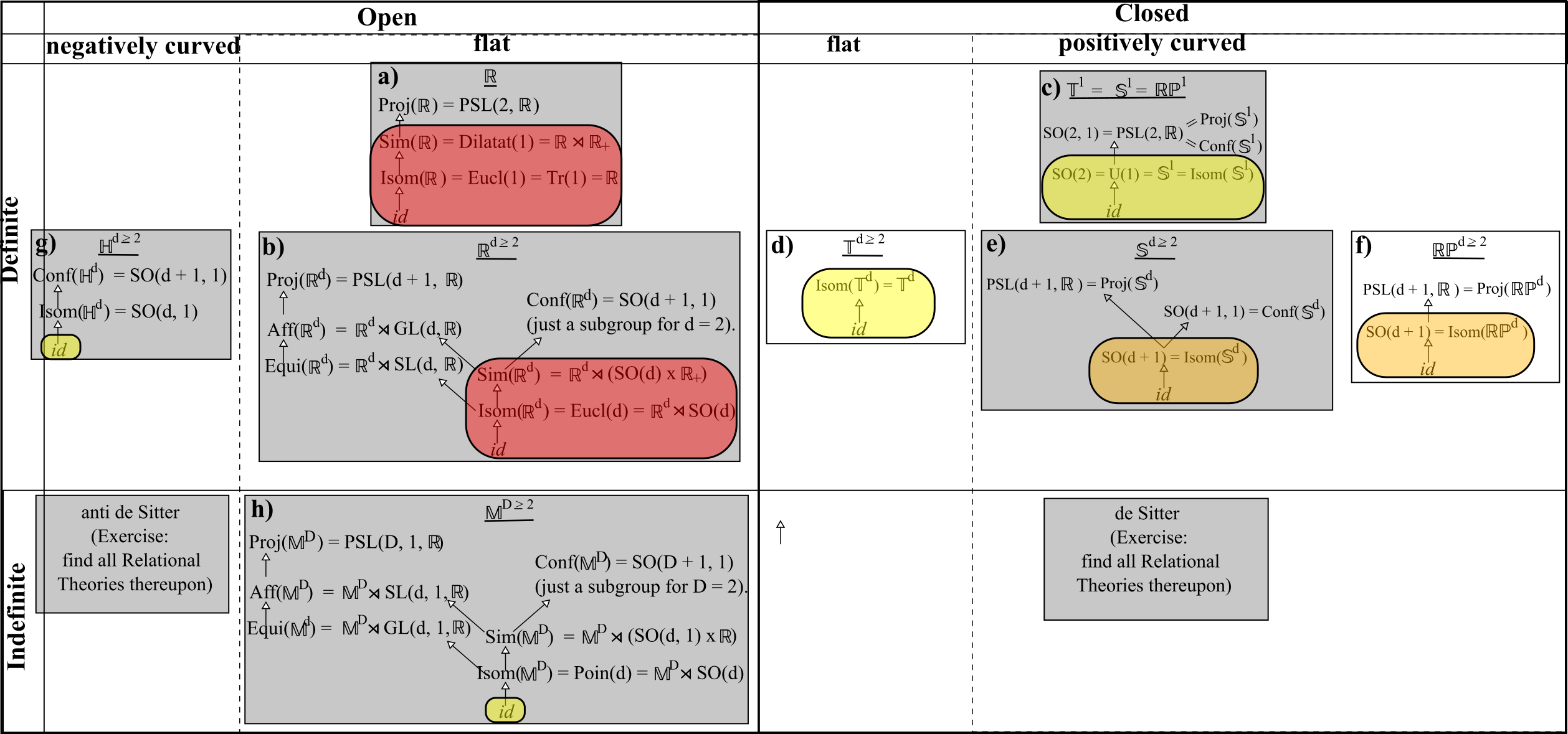}
\caption[Text der im Bilderverzeichnis auftaucht]{\footnotesize{Poset of geometrically significant automorphism groups yielding relational theories, 
now encircling which cases are protected by the current Article's 'quotient by a compact Lie group' criterion, for  
a) $\mathbb{R}$, 
b) $\mathbb{R}^d$ for $d \geq 2$,
c) $\mathbb{T}^1 = \mathbb{S}^1 = \mathbb{RP}^1$, 
d) $\mathbb{T}^d$, 
e) $\mathbb{S}^d$, 
f) $\mathbb{RP}^d$ (each for $d \geq 2$),
g) $\mathbb{H}^2$,  
h) $\mathbb{M}^D$ for $D \geq 2$.   
Kendall's Shape Theory and its partner Leibnizian Shape-and-Scale Theory are in the protected groupings shaded in red. 
Other nontrivial groupings protected by this criterion are shaded in orange. 
Surviving trivial Relational Theories are in yellow.
All of curvature, non-maximality, topological identification reduce number of tractable Relational Theories in both Article I and Article II's manner. 
Carrier space indefiniteness, moreover, which did not impinge on diversity of Shape Theories in Article I, places all nontrivial spacetime Relational Theories 
outside of this protective compactness criterion. }} 
\label{Cpct1}\end{figure} } 
%FFFFFFFFFFFFFFFFFFFFFFFFFFFFFFFFFFFFFFFFFFFFFFFFFFFFFFFFFFFFFFFFFFFFFFFFFFFFFFFFFFFFFFFFFFFFFFFFFFFFFFFFFFFFFFFFFFFFFFFFFFFFFFFFFFFFFFFFFFFFFFFFFFFFFFFFFFFFFFFFFFFFFFFFFFFFFFFFFFFFFFFFFF

%===================================================================================================================================================================================
%===================================================================================================================================================================================
\section{Comparison with Kendall--Barden--Carne--Le treatment of shape space topology}
%===================================================================================================================================================================================
%===================================================================================================================================================================================

\n They \cite{Kendall} looked at the similarity shapes case in detail. 

\m 

\n Firstly, the corresponding shape spaces possess a metric that is compatible with their topology. 
So, using the corresponding balls to `house off', Hausdorffness is guaranteed.

\m 

\n They furthermore showed that these shape spaces can be modelled as finite CW spaces -- closure-finiteness spaces equipped with the weak topology -- 
(see \cite{Janich, Hatcher} for pedagogical introductions to such and \c{Dold} for product and quotient theorems for these).
Being CW moreover implies local contractability, compactness and paracompactness. 
Compactness implying local compactness, we thus arrive at Kendall's similarity shape spaces being LCHP.  
This renders Pflaum's approach \cite{P00, Pflaum} to stratified LCHP spaces\footnote{Pflaum's work post-dates Kendall et al's book review by 1 to 2 years.} 
applicable to furthering Kendall's Shape Theory.    

\m 

\n My own way of looking at these shape spaces -- as quotients by compact groups: Sec 8 -- gives that they are LCHS.

\m 

\n By (\ref{LCHSP}), I recover most of Kendall et al's topological-space level conclusions.  
From this position, moreover, Kreck's and Sniatycki's approaches to LCHS stratified manifolds -- 1 decade more modern than Pflaum's -- become applicable.

%===================================================================================================================================================================================
%===================================================================================================================================================================================
\section{Mumford's criterion and it relational range of applicability}
%===================================================================================================================================================================================
%===================================================================================================================================================================================

\n  Groisser and Tagare \cite{GT09} moreover describe the phenomenon impossibilating affine shape space's Hausdorffness (and accessibility), as follows.  

\m 

\n ``{\it Suppose a group $\lFrg$ acts on a topological space $X$, that $p \in X$ is a point whose orbit ${\cal O}_p$ is not closed in $X$, 
                                                     and that $q \in X$ lies in the closure of ${\cal O}_p$ but not in ${\cal O}_p$ itself.  
Then, letting $\w{p}$, $\w{q}$ denote the images of $p$, $q$ in the quotient $X/\lFrg$, 
every open neighborhood of $\w{q}$ contains $\w{p}$.}"  

\m 

\n We now clarify that this is a `Mumfordian' phenomenon, with reference to Mumford's 1965 approach here (see \cite{MFK94} for an updated review).  
He considers 
\be
\frac{X}{\lFrg} \m \mbox{ for $X$ now in general a variety} \m .    
\ee 
Some orbits are then `discarded for not being stable', making this a more advanced excision strategy.  
The criterion for what is unstable and so to be discarded in this approach is the above-quoted one. 

\m 

\n We note that this is not stated to {\sl always} occur upon quotienting out nocompact groups, or even in the presence of nonproper actions; 
see the next subsection for simple counterexamples.  

\m 

\n As regards the applicability of such an excision, on the one hand, in  Shape Statistics, Image Analysis and Computer Vision, 
this holds up because only approximate data are ever observed. 
So one can contend that, e.g.\, location data never contains any exact collinearities in the first place. 
By which the corresponding stratum of collinear configurations can be removed from the modelling. 

\m 

\n On the other hand, in Geometry collinear configurations and other nonprincipal strata shapes have an exact existence. 
In Classical Dynamics, moreover, dynamical trajectories necessitate handling passage through lower-strata states. 
Article III comments further on details and significance of these matters.

%===================================================================================================================================================================
%===================================================================================================================================================================
\section{Conclusion}
%===================================================================================================================================================================
%===================================================================================================================================================================

\n \cite{A-Generic} placed a great bound on the scope of Relational Theory (Shape and Shape-and Scale Theory) at the differential-geometric level.
This follows from manifolds possessing generalized Killing vectors being nongeneric.  
It is a useful observation given recent diversification posited and at least partly solvable models in this subject 
\cite{Kendall, Sparr, MP03, MP05, FileR, Quad-I, Bhatta, AMech, PE16, I-II-III, ACirc, ASphe, ATorus, Aff}.  

\m 

\n The current Article moreover places places a second great bound, now at the level of topological spaces: that compactness guarantees technical tractability.
From this point of view, Kendall's Shape Theory's own success is tied to two instances of compactness: that of the preshape sphere 
                                                                                      and, especially, that of the rotation group that is quotiented out from the preshape sphere. 

\m 

\n Within the nongeneric set of geometrically-equipped differentiable manifold carrier spaces that do possess nontrivially Relational Theories, 
the corresponding relational spaces are in general stratified manifolds rather than manifolds. 
Representability of a relational space as a quotient by the continuous action of a compact Lie group moreover translates to that relational space's good analytic behaviour. 
It is in particular Hausdorff, second-countable and locally-compact (LCHS), which is sufficient to imply paracompactness as well.  
A further case involves isolated bad points, as exemplified by the usual Euclidean Shape-and-Scale Theory's maximal coincidence-or-collision. 
Kendall's own Shape Theory avoids this by being of the form 
\be 
\frac{\mbox{compact manifold}}{\mbox{compact Lie group}}  \es  \frac{\FrM_{\sC}}{\lFrg_{\sC}}  \m ,  
\ee 
as realized by 
\be 
\frac{\mbox{compact preshape sphere}}{\mbox{compact rotation group}}  \es  \frac{\mathbb{S}^{n \, d - 1}}{SO(d)}  \m .    
\ee
When the quotiented-out group is not compact, however, widespread bad analytic behaviour -- more specifically merely-Kolmogorov rather than Hausdorff separation -- 
has been sighted. 
This occurs for instance in flat-space Affine and Projective Shape Theory \cite{GT09, KKH16}, 
which are desirable theories from the point of view of Image Analysis and Computer Vision. 
We note that these examples moreover lie {\sl outside} of the remit of our protective compactness criteria. 

\m 

\n We subsequently tabulated which Relational Theories put forward to date lie within our compactness criteria, and which lie outside: see Fig \ref{Cpct1}.  
This unfortunately only guarantees good beaviour of a fairly small subset of Relational Theories in addition to Kendall's own highly sucessful prototype.
Indeed, we suggest here that Kendall's Shape Theory's success is in good part due to its complying with our two compactness selection criteria for technical tractability.  

\m 

\n We also presented various further arguments against excising strata or unfolding strata, by which one is left accepting strata. 
In this case, fibre bundle methods also cease to suffice, due to different strata in general having distinct objects attached to them rather than homogeneous fibres.  
LCHS stratified manifolds moreover lie within the remit of \'{S}niatycki's \cite{SniBook}, Kreck's \cite{Kreck} and Pflaum's \cite{P00, Pflaum} methodology.
Part of the first of these approaches, involving {\sl subcartesian differential spaces}, is outlined in the current Article as a further source of results for 
analytically well-behaved stratified relational spaces. 
The other two are based on {\sl (stratified manifold, sheaf) pairs}, and were briefly considered in the relational space context in \cite{ABook, A-Generic}.  

\m 

\n Note the LCHP to LCHS upgrade, by which one can pass from using Pflaum's approach to stratified manifolds to Kreck's and/or \'{S}niatycki's.

\m 

\n Moreover, since the case of Relational Theory that do not lie within the remit of our protective compactness selection critera are well-motivated, 
with particular interest in the Affine and Projective Shape Theories from Image Analysis and Computer Vision points of view. 
On these grounds, further attempts to understand, or get around, merely-Kolmogorov spaces in these cases are likely to be attempted.  

\m  

\n Let us end by pointing to two further research frontiers. 

\m 

\n{\bf Research Frontier 1} We have not looked at explicit examples of supersymmetric shapes, i.e.\ whether our protective compactness selection criteria 
extend to super-shape constellation-space supermanifolds quotiented by 

\n supergeometrically-significant automorphism supergroups. 

\m 

\n{\bf Research Frontier 2} Article III's Comparative Background Independence \cite{APoT, ABook, PE-1, DO-1} extends 
moreover beyond Article I and II's remit of Differential-Geometrical Background Independence to Topological Background Independence and beyond.  
In this wider setting, the automorphism groups one is quotienting out cease to be not only geometrically-significant but also to no longer be Lie groups (compact or otherwise). 

\m

\n{\bf Acknowledgments} I thank Professor Chris Isham for discussions.
I also pay my respects to Professor Graham Allan, 
who first brought subtleties with quotient spaces to my attention when I was an undergraduate and subsequently passed away,  
and also thank Professor Timothy Gowers for some inspirational lectures in elementary topology. 
I also thank Professors Don Page, Jeremy Butterfield, Enrique Alvarez, Reza Tavakol and Malcolm MacCallum for support with my career, 
and both the Isaac Newton Institute and the Cambridge Algebraic Geometry group for some stimulating seminars and useful discussions.  
I finally dedicate this Article to my most loyal friend; I appreciate your style.

%=====================================================BIBLIOGRAPHY==========================================================================================================================

\end{document}